\documentclass[aps,pra,groupedaddress,tightenlines, twocolumn]{revtex4}
\usepackage{graphicx}
\usepackage{amsmath}
\usepackage[english]{babel}

\begin{document}

\title{Dynamics of a two-level system under the simultaneous influence of a spin bath and a boson bath}

\author{Ning Wu and Yang Zhao\footnote{Electronic address:~\url{YZhao@ntu.edu.sg}}}

\affiliation{\it School of Materials Science and Engineering,
Nanyang Technological University, Singapore 639798}
\date{\today}

\begin{abstract}
We study dynamics of a two-level system coupled simultaneously to a pair of dissimilar reservoirs, namely, a spin bath and a boson bath, which are connected via finite interbath coupling.
It is found that the steady-state energy transfer in the two-level system increases with its coupling to the spin bath while optimal transfer occurs
at intermediate coupling in the transient process.
If the two-level system is strongly coupled to the spin bath, the population transfer is unidirectional barring minor population oscillations of minute amplitudes.
If the spin bath is viewed as an atomic ensemble, robust generation of macroscopic superposition states exists against parameter variations of the two-level system and the boson bath.
\end{abstract}

\maketitle


\section{Introduction}

With a consequence that typically includes decoherence and dissipation \cite{Breuer}, any realistic quantum system is inevitably coupled to its surrounding environment, which is believed to play a detrimental role in processes such as resonant energy transfer, quantum information processing, and spin manipulation in semiconductors. Great efforts have been devoted to understanding the decoherence process in solid-state spin nanodevices, one of the most promising candidates for quantum information processing and computation\cite{Loss-1998,Loss-1999}. A dominant contribution to quantum decoherence in solid-state spin nanostructures arises from the nuclear spins. Several models have been proposed to study the properties of a two-level system (TLS) interacting with a spin environment \cite{spinbath,Stamp-2000,Zhang-2007}, for which a commonly used setup is a preferred central TLS coupled homogeneously to a bath of surrounding spins with no intrabath interactions \cite{Bose-2002,Breuer-2004,PRL2012,Sarma}. Coupling of a central TLS with a spin bath can in general lead to non-Markovian behavior, which, as suggested in Ref.~\cite{PRL2012}, plays an important role in energy transport in biological systems. On the other hand, the spin-boson model \cite{Leggett}, an extensively studied system in the context of quantum decoherence, has seen a large variety of applications in fields ranging from quantum information processing\cite{Uhrig} to light-harvesting systems \cite{Mckenzie1,Mckenzie2,JPCL}.

In this work, both types of environments, namely, a spin bath and a boson bath, are included in assessing their effects on the central TLS. We first focus on its polarization dynamics of the TLS. As illustrated in Fig.~1, our model can be viewed as an extension of the conventional spin-boson model\cite{Leggett} to include an additional spin bath \cite{PRL2012,Breuer-2004}, and is capable to capture the interplay of the two baths. The interaction between the TLS and the spin bath is assumed to be of the Ising type, and correlations between the two baths are taken into account via linear coupling. By viewing the spin bath as a system of interest, we also investigate the influence of the TLS and the boson bath on the macroscopic superposition generation in the spin bath, which is prepared, for example, with all spins in the bath oriented along the $+\hat{x}$ direction). In this context, Ref.~\cite{Bose-2002} examined the TLS-induced correlation and entanglement in the spin bath, and Ref.~\cite{kurizki} studied the macroscopic superposition generation of an ensemble of atoms assisted by a linearly coupled boson bath. In this work we investigate the effect of the TLS on a tripartite system similar to that in Ref.~\cite{kurizki}. Robust generation is found to exist against parameter variations of the TLS and the boson bath.

The rest of paper is structured as follows. In Sec.~II, we introduce the model and methodology employed in this work. In Sec.~III, results on polarization dynamics of the TLS are described in great detail. In Sec.~IV, we discuss the macroscopic quantum-superposition states generation in the spin bath. Conclusions are drawn in Sec.~V.

\begin{figure}
\includegraphics[scale=0.56, angle=0]{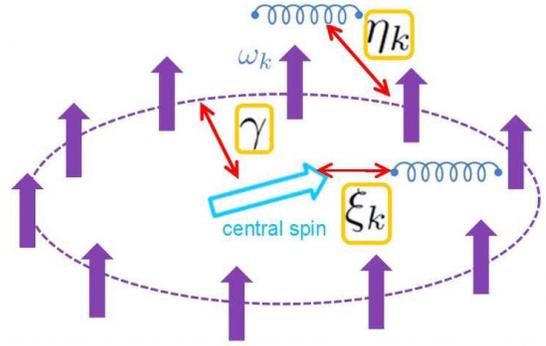}
\caption{Schematics of a TLS (blue, central spin) coupled to a spin bath and a boson bath. The spin bath, which has no intrabath interactions, interacts uniformly via Ising-like coupling to the TLS. Both the TLS and the spin bath are coupled to the boson bath via conventional linear spin-boson coupling.}
\end{figure}

\section{The model}

Our model Hamiltonian $H$ is composed of four parts:
\begin{eqnarray}\label{H}
H&=&H_{\rm T}+H_{\rm S}+H_{\rm I}+H_{\rm B},\\
H_{\rm T}&=&\frac{\varepsilon}{2}\sigma_z+J\sigma_x,~H_{\rm S}=\alpha L_z+\gamma L_z\sigma_z,\nonumber\\
H_{\rm I}&=&\sum_k (L_z\eta_k+\frac{\sigma_z}{2}\xi_k)(b_k+b^\dag_k),~H_{\rm B}=\sum_k\omega_kb^\dag_kb_k.\nonumber
\end{eqnarray}
Here, $H_{\rm T}$ describes the TLS (a two-level system) with energy level spacing $\varepsilon$ (with $\sigma_{x,z}$ being the Pauli matrices), $J$ is the transfer integral between the two states. $H_{\rm B}$ describes the boson bath with annihilation (creation) operator $b_k$ ($b_k^\dagger$) of the bath mode with frequency $\omega_k$. $H_{\rm S}$ is the Hamiltonian for the spin bath including its interaction with the TLS, $\alpha$ is an energy proportionality constant, and $L_z=\sum^N_{j=1}\sigma^z_j/2$ is the $z$-component of the total spin operator $\vec{L}$, describing a bath of $N$ spin-1/2 noninteracting spins ($N$ is an even number). $H_{\rm I}$ captures interactions of the boson bath with the TLS and the spin bath: The TLS interacts with the spin bath via the TLS-spin-bath (T-S) coupling $\gamma$ in addition to its coupling to the conventional boson bath through the TLS-boson-bath (T-B) coupling $\xi_k$; There are also interactions between the spin bath and the boson bath (the S-B coupling) which are denoted by $\eta_k$ in our model.
The effects of the boson bath are fully captured by the spectral densities. Cubic spectral densities are assumed here:
\begin{eqnarray}
J_{\rm TT}(\omega)&=&\sum_k\xi^2_k\delta(\omega-\omega_k)=\kappa_1\omega^{-2}_{\rm ph}\omega^3 e^{-{\omega}/{\omega_c}},\label{SF1}\\
J_{\rm SS}(\omega)&=&\sum_k\eta^2_k\delta(\omega-\omega_k)=\kappa_3\omega^{-2}_{\rm ph}\omega^3 e^{-{\omega}/{\omega_c}},\label{SF2}\\
J_{\rm TS}(\omega)&=&\sum_k\xi_k\eta_k\delta(\omega-\omega_k)=\kappa_2\omega^{-2}_{\rm ph}\omega^3 e^{-{\omega}/{\omega_c}},\label{SF3}
\end{eqnarray}
where $\omega_c$ is the cutoff frequency and $\omega_{\rm ph}$ is the characteristic phonon frequency and will be used as the energy unit.
$J_{\rm TT}(\omega)$ in Eq.~(\ref{SF1}) is the conventional spin-boson spectral density function
with $\kappa_1$ the coupling strength quantifying the TLS coupling to the boson bath, while $J_{\rm SS}(\omega)$ in Eq.~(\ref{SF2}) denotes
the corresponding spectral density for the interactions between the spin-bath and the boson bath
with $\kappa_3$ the coupling strength. $J_{\rm TS}(\omega)$ in Eq.~(\ref{SF3})
describes the hybridization of $J_{\rm TT}(\omega)$ and $J_{\rm SS}(\omega)$. Note that $\kappa_2$ can be varied independently of $\kappa_1$ and $\kappa_3$ through sign changes of $\xi_k's$ or $\eta_k's$.
As we will see below, the spectral density $J_{\rm SS}(\omega)$ does not effect the reduced dynamics of the TLS within the framework of the polaron master equation approach used here, and the bath-bath correlation is controlled by $J_{\rm TS}(\omega)$ alone. However, $J_{\rm SS}$ will be present in the reduced dynamics of the spin bath.

Note that $H$ commutes with $ L^2=\sum_iL^2_i$ ($i=x,y,z$) and $L_z$, so we can restrict ourselves in each $(l,m)$-sector, where $l$ is the total spin which runs from 0 to $\frac{N}{2}$ for even $N$, and $m$ is the eigenvalue of $L_z$.
When the T-B coupling is absent, i.e., $\xi_k=0$, $H$ can be divided into two commuting parts: $H=H_1+H_2$ with $H_1=H_{\rm T}+H_{\rm S}$, and $H_2=L_z\sum_k\eta_k(b_k+b^\dag_k)+H_{\rm B}$. Noting that $[H_1,H_2]=0$, the boson bath has no effect on the reduced dynamics of the TLS and we recover the results of Ref.~\cite{PRL2012}, and dynamics of $H_1$ and $H_2$ can both be obtained analytically \cite{kurizki,PRL2012}. For finite T-B coupling, the dynamics of Eq.~(\ref{H}) can not be solved exactly.
In the next section, we will adopt the recently proposed time-convolutionless (TCL) polaron master equation approach\cite{Jang} to study the reduced dynamics of the TLS.
The approach takes into account simultaneously correlated initial conditions and effects of strong coupling and non-Markovian baths.

\section{Reduced dynamics of the TLS}

\subsection{The polaron transformation}
 The polaron canonical transformation is generated by
 \begin{eqnarray}\label{PT}
 S&=&L_zB_1+\frac{\sigma_z}{2}B_2,\nonumber\\
 B_{1 }&=&\sum_k\frac{\eta_k }{\omega_k}(b^\dag_k-b_k),~B_{2}=\sum_k\frac{ \xi_k }{\omega_k}(b^\dag_k-k_k),
  \end{eqnarray}
resulting in the transformed Hamiltonian $\tilde{H}$ given by
\begin{eqnarray}\label{Htilde}
\tilde{H}&=&e^{S}He^{-S}=\tilde{H}_0+\tilde{H_1}+H_{\rm B},\\
\tilde{H}_0&=&\frac{\varepsilon}{2}\sigma_z+\tilde{J}\sigma_x+\alpha L_z+\tilde{\gamma}L_z\sigma_z-L^2_z\eta,\label{eq5}\\
\tilde{H}_1&=&J[\sigma_x(\cosh B_2-\Theta)+i\sigma_y\sinh B_2],
\end{eqnarray}
where $\tilde{\gamma}=\gamma-\sum_k{\eta_k\xi_k}/{\omega_k}$, $\Theta=\langle\cosh B_2\rangle$, $\tilde{J}=J\Theta$, and $\eta=\sum_k{\eta^2_k}/{\omega_k}$. Here a finite transfer integral $J$ is always assumed, and the average $\langle ... \rangle$ is taken over the boson bath in thermal equilibrium at temperature $T$. Note that only the T-B coupling ($\xi_k$) enters the expression of the renormalized transfer integral $\tilde{J}$, while both the T-B and the S-B coupling ($\xi_k$ and $\eta_k$)
show up in the expression of the effective T-S coupling $\tilde{\gamma}$. Furthermore, the interbath interaction will induce
a nonlinear term $-L^2_z\eta$.

The technique of polaron transformation, helpful to identify Hamiltonian terms that remain small beyond the weak exciton-phonon coupling regime, was employed earlier by Munn and Silbey\cite{MS1,zhao94}, for example,
to study transport properties of molecular crystals. In recent years, this approach has been used in a variety of disciplines such as physical chemistry (to study excitation energy transfer in light-harvesting systems) \cite{Jang,Nazir2011,JCP2012}, quantum optics \cite{Hughes}, and biophysics \cite{DNA}. Note that by construction, $\langle\tilde{H}_1\rangle=0$.
Below the renormalized hopping term, $\tilde{H}_1$, will be treated perturbatively, in an approximation that is believed to be valid even in the intermediate or strong system-bath coupling regime, especially for fast baths with $\omega_c\ge 2J$.

To apply the TCL master equation to Eqs.~(4)-(6), it is convenient to work in the interaction picture with respect to $\tilde{H}_0+H_{\rm B}$. We can diagonalize $\tilde{H}_0$ in each $(l,m)$-sector:
\begin{eqnarray}
\tilde{H}_0&=&\sum^{\frac{N}{2}}_{l=0}\sum^l_{m=-l}\tilde{H}_0^{(lm)}|l,m\rangle\langle l,m|,\nonumber\\
\tilde{H}_0^{(lm)}&=&\frac{\varepsilon}{2}\sigma_z+\tilde{J}\sigma_x+\alpha m+\tilde{\gamma}m\sigma_z-m^2\eta,
\end{eqnarray}
where $|l,m\rangle$ is the eigenstate of the spin bath. The two eigenstates of $\tilde{H}_0^{(lm)}$ read
\begin{eqnarray}
|\phi^+_m\rangle&=&\cos\frac{\theta_m}{2}|1\rangle+\sin\frac{\theta_m}{2}|\bar{1}\rangle,\nonumber\\
|\phi^-_m\rangle&=&-\sin\frac{\theta_m}{2}|1\rangle+\cos\frac{\theta_m}{2}|\bar{1}\rangle.
\end{eqnarray}
Here, $|1\rangle$ and $|\bar{1}\rangle$ are the two eigenstates of $\sigma_z$ with eigenvalues $\pm1$. The corresponding eigenenergies are
\begin{eqnarray}
E_{\pm}(m)=(\alpha m-\eta m^2)\pm\frac{\epsilon_m}{2},
\end{eqnarray}
 where $\epsilon_m=E_+(m)-E_-(m)=(4\tilde{J}^2+{\tilde{\varepsilon}_m}^2)^{1/2}$, with $\tilde{\varepsilon}_m=\varepsilon+2\tilde{\gamma}m$. The rotation angle $\theta_m$ is determined by $\tan\theta_m= 2\tilde{J}/\tilde{\varepsilon}_m$. By further defining the pseudo-Pauli matrices in the $|\phi^\pm_m\rangle$ basis:
 \begin{eqnarray}
\tau^z_m&=&|\phi^+_m\rangle\langle\phi^+_m|-|\phi^-_m\rangle\langle\phi^-_m|,\nonumber\\
\tau^+_m&=&|\phi^+_m\rangle\langle\phi^-_m|, \tau^-_m=(\tau^+_m)^\dag,
  \end{eqnarray}
  we can get $\tilde{H}_1$ in the interaction picture
 \begin{eqnarray}\label{H1I}
\tilde{H}_{1,I}(t)=J\tilde{\sigma}_+(t)D(t)+{\rm H.c.},
\end{eqnarray}
where H.c.~stands for Hermitian conjugate, and
   \begin{eqnarray}
   \tilde{\sigma}_+(t)&=&\sum^{\frac{N}{2}}_{l=0}\sum^l_{m=-l}K^k_m(t)\tau^k_{m}|l,m\rangle\langle l,m|,\\
   D(t)&=&e^{e^{iH_{\rm B}t}B_2e^{-iH_{\rm B}t}}-\Theta,\nonumber\\
K^x_m(t)&=&\frac{1}{4}[(1+C_m)e^{i\epsilon_mt}-(1-C_m)e^{-i\epsilon_mt}],\nonumber\\
 K^y_m(t)&=&\frac{i}{4}[(1+C_m)e^{i\epsilon_mt}+(1-C_m)e^{-i\epsilon_mt}],\nonumber\\
 K^z_m(t)&=&\frac{S_m}{2}.\nonumber
 \end{eqnarray}
Here we have defined $C_m\equiv\cos\theta_m,~S_m\equiv\sin\theta_m$.

\subsection{Observables and initial conditions}

Before applying the projection operator technique\cite{Breuer} to trace out the boson degrees of freedom, we first build a connection between density matrices and physical observables.
Let $\rho(t)$ denote the Schr\"{o}dinger picture total density matrix of the entire system, and $\sigma(t)$, the reduced density matrix after tracing $\rho(t)$ over the boson degrees of freedom only. Their counterparts in the polaron frame are labeled as $\tilde{\rho}(t)$ and $\tilde{\sigma}(t)$
Let $\rho(t)$, $\sigma(t)$ and $\theta(t)$ denote the Schr\"{o}dinger picture density operators of the entire system, the sum of the TLS and the spin bath, and the TLS itself, respectively, while their counterparts in the polaron frame are labeled as  $\tilde{\rho}(t),\tilde{\sigma}(t)$ and $\tilde{\theta}(t)$.

We are interested in real-time dynamics of the observables
\begin{eqnarray}\label{Si}
\langle\sigma_i\rangle&=&{\rm Tr}_{\rm T+S+B}[\rho(t)\sigma_i],~i=x,y,z,
\end{eqnarray}
where the sub-indices of $\rm{Tr}$,  T, S, and B, indicate traces over degrees of freedom of the TLS (T), the spin bath (S), and the boson bath (B), respectively. In the projection operator approach, the polaron transformed total density matrix in the interaction picture $\tilde{\rho}_{\rm I}(t)$ can be divided into the relevant part $\mathcal{P}\tilde{\rho}_{\rm I}(t)={\rm Tr}_{\rm B}[\tilde{\rho}_{\rm I}(t)]\rho_{\rm B}=\tilde{\sigma}_{\rm I}\rho_{\rm B}$ and the irrelevant part $\mathcal{Q}\tilde{\rho}_{\rm I}(t)$, where the super-operator $\mathcal{P}$ is defined by $\mathcal{P}(\cdot)=\rho_{\rm B}\otimes {\rm Tr}_{\rm B}(\cdot)$ and $\mathcal{Q}=1-\mathcal{P}$. Correspondingly, the expectation values $\langle\sigma_i\rangle$ can be written as $\langle\sigma_i\rangle=\langle\sigma_i\rangle_P+\langle\sigma_i\rangle_Q$, i.e., the summation of the relevant part $\langle\sigma_i\rangle_P$ and irrelevant part $\langle\sigma_i\rangle_Q$. In general, the TCL master equation only yields dynamics for the relevant part. However, thanks to the fact that $[\sigma_z,S]=0$, the irrelevant contribution of $\langle\sigma_z\rangle$ vanishes: $\langle\sigma_z\rangle_Q={\rm Tr}_{\rm T+S+B}[\mathcal{Q}\tilde{\rho}_{\rm I}(t)e^{i\tilde{H}_0t}\sigma_ze^{-i\tilde{H}_0t}]=0$. It is easily seen that
\begin{eqnarray}\label{sigma}
\left(
  \begin{array}{c}
    \langle\sigma_x\rangle_P \\
    \langle\sigma_y\rangle_P \\
    \langle\sigma_z\rangle_P \\
  \end{array}
\right)=\left(
          \begin{array}{c}
            \Theta {\rm Tr}_{\rm T+S}[\tilde{\sigma}(t)\sigma_x] \\
            \Theta {\rm Tr}_{\rm T+S}[\tilde{\sigma}(t)\sigma_y]\\
            {\rm Tr}_{\rm T+S}[\tilde{\sigma}(t)\sigma_z] \\
          \end{array}
        \right).
\end{eqnarray}
Note that in general $\langle\sigma_x\rangle_Q$ and $\langle\sigma_y\rangle_Q$ do not vanish. But they do not participate in the closed system of equations of motion for $\{\langle\sigma_i\rangle_P\}$ neither. Since the total angular momentum $l$ of the spin bath is conserved, it is convenient to introduce the $l$-independent quantities
\begin{eqnarray}\label{alpha}
\alpha^i_{m}(t)&=&{\rm Tr}_{\rm T}[\langle l,m|\tilde{\sigma}(t)|l,m\rangle\sigma_i],~i=x,y,z,\nonumber\\
\alpha^e_{m}(t)&=&{\rm Tr}_{\rm T}[\langle l,m|\tilde{\sigma}(t)|l,m\rangle].
\end{eqnarray}
Using the trace formula over the spin-bath
\begin{eqnarray}\label{trace}
{\rm Tr}_{\rm S}(\cdot)=\sum^{\frac{N}{2}}_{l=0}\sum^l_{m=-l}\nu(l,\frac{N}{2})\langle l,m|\cdot|l,m\rangle,
\end{eqnarray}
where $\nu(l,\frac{N}{2})\equiv C^{l+N/2}_N-C^{l+1+N/2}_N$ denotes degeneracy of the spin bath \cite{PRL2012,Molmer-2002,Pet-2006}, it can be seen that
\begin{eqnarray}
&&{\rm Tr}_{\rm T+S}[\tilde{\sigma}(t)\sigma_i]=\sum_{lm}\nu(l,\frac{N}{2})\alpha^i_{m}(t),i=x,y,z\nonumber\\
&&\sum_{lm}\nu(l,\frac{N}{2})\alpha^e_{m}(t)=1.
\end{eqnarray}
A separable initial state is assumed in our model:
\begin{eqnarray}\label{IS}
\rho(0)=\rho_{\rm T}\otimes\rho_{\rm S}\otimes\rho_{\rm B}=|\bar{1}\rangle\langle\bar{1}|\otimes\frac{e^{-\beta\alpha L_z}}{Z_{\rm S}}\otimes \frac{e^{-\beta H_{\rm B}}}{Z_{\rm B}},
\end{eqnarray}
where $\beta=1/k_{\rm B}T$ is the inverse temperature, and the two baths are in thermal equilibrium with partition functions for the spin and boson baths given by $Z_{\rm S}$ and $Z_{\rm B}$, respectively. Although separable in the original frame, the two baths are entangled after being transformed into the polaron frame:
 \begin{eqnarray}\label{rho0}
\tilde{\rho}(0)=|\bar{1}\rangle\langle\bar{1}|\sum_{lm}|l,m\rangle\langle l,m|\tilde{\rho}^m_{\rm B}\frac{ {e^{-\beta\alpha m}}}{{Z_{\rm S}}},
\end{eqnarray}
where $\tilde{\rho}^m_{\rm B}=e^{mB_1-\frac{1}{2}B_2}\rho_{\rm B}e^{-mB_1+\frac{1}{2}B_2}$. Correspondingly, the initial values of $\alpha's$ read:
 \begin{eqnarray}\label{initialalpha}
\alpha^x_{m}(0)=\alpha^y_{m}(0)=0, \alpha^z_{m}(0)=-\alpha^e_{m}(0)=-\frac{e^{-\beta\alpha m}} {Z_{\rm S}}.
\end{eqnarray}

\subsection{Polaron master equation for the TLS}

Now we can apply the TCL master equation  to $\tilde{\sigma}_{\rm I}(t)$. The standard projection operator technique leads to the following master equation for $\tilde{\sigma}_I(t)$\cite{Breuer,Jang}
\begin{eqnarray}
\frac{d\tilde{\sigma}_I(t)}{dt}&=&\int^t_0ds\rm{Tr}_{\rm B}\{\tilde{\mathcal{L}}_{1,I}(t)\tilde{\mathcal{L}}_{1,I}(s)\tilde{\sigma}_I(t)\rho_B\}\nonumber\\
&&+\rm{Tr}_{\rm B}\{\tilde{\mathcal{L}}_{1,I}(t)\mathcal{Q}\tilde{\rho}(0)\}\nonumber\\
&&+\int^t_0ds\rm{Tr}_{\rm B}\{\tilde{\mathcal{L}}_{1,I}(t)\tilde{\mathcal{L}}_{1,I}(s)\mathcal{Q}\tilde{\rho}(0)\},
\end{eqnarray}
where $\tilde{\mathcal{L}}_{1,I}(t)(\cdot)=-i[\tilde{H}_{1,I}(t),\cdot]$. The last two terms in the above equation are the first and second order inhomogeneous terms, respectively. It is obvious that $\dot{\alpha}^e_{lm}(t)=0$, so that $\alpha^e_{m}={e^{-\beta\alpha m}}/{Z_{\rm S}}$ is a constant. Transforming back to the Schr\"{o}dinger picture and using Eqs.~(\ref{H1I}),(\ref{alpha}) and (\ref{rho0}), we find after a tedious but straightforward calculation the following set of Bloch equations, which is obeyed by the vector $\vec{\alpha}_m(t)=[\alpha^x_m(t),\alpha^y_m(t),\alpha^z_m(t)]^T$:
\begin{eqnarray}\label{Bloch}
\dot{\vec{\alpha}}_m(t)=M(t) \vec{\alpha}_m(t)+\vec{R}_m(t),
\end{eqnarray}
with
\begin{eqnarray}\label{xyz}
\vec{R}_m(t)&=&\vec{R}^{(e)}_m(t)+\vec{R}^{(1)}_m(t)+\vec{R}^{(2)}_m(t),\\
M(t)&=&\left(
          \begin{array}{ccc}
            -G^{1-}_{my} & G^{1-}_{mx}-\tilde{\varepsilon}_m & 0 \\
            G^{2+}_{my}+\tilde{\varepsilon}_m & -G^{2+}_{mx} & -2\tilde{J}\\
            G^{2+}_{mz} & G^{1-}_{mz}+2\tilde{J} &-(G^{2+}_{mx}+G^{1-}_{my}) \\
                  \end{array}
        \right),\nonumber
\end{eqnarray}
where $\vec{R}^{(\kappa)}_m(t)=[R^{\kappa}_{mx}(t),R^{\kappa}_{my}(t),R^{\kappa}_{mz}(t)]^T$ for $\kappa=e,1$ and $2$. The homogeneous rates $G^{\xi+}_{mi}=J^2(\gamma^\xi_{mi}+\gamma^{\xi*}_{mi})$ and $G^{\xi-}_{mi}=i J^2(\gamma^\xi_{mi}-\gamma^{\xi*}_{mi})$($\xi=1,2;~i=x,y,z$) appearing in the Bloch matrix $M(t)$ are combinations of
\begin{eqnarray}\label{Hcf}
\gamma^1_{mi}(t)&=&\Theta^2\int^t_0ds \tilde{K}^{i-}_m(-s)[e^{-\phi(s)}-e^{\phi(s)}],\nonumber\\
\gamma^2_{mi}(t)&=&\Theta^2\int^t_0ds\tilde{K}^{i+}_m(-s)[e^{-\phi(s)}+e^{\phi(s)}-2],
\end{eqnarray}
where the bath correlation function $\phi(s)$ has the form
\begin{eqnarray}\label{CF}
\phi(s)=\sum_k\left(\frac{\xi_k} {\omega_k}\right)^2\left(\cos \omega_k s \coth \frac{\beta\omega_k}{2} -i\sin \omega_k s\right),
\end{eqnarray}
and $\tilde{K}^{i\pm}_m(s)\equiv\tilde{K}^{i}_m(s)\pm\tilde{K}^{i*}_m(s)$ with $\tilde{K}^{i}_m(s)$ given explicitly by
 \begin{eqnarray}\label{Km}
\tilde{K}^x_m(s)&=&C_mK^x_m(s)+S_mK^z_m(s),\nonumber\\
\tilde{K}^y_m(s)&=&K^y_m(s),\nonumber\\
\tilde{K}^z_m(s)&=&-S_mK^x_m(s)+C_mK^z_m(s).
\end{eqnarray}
The inhomogeneous part $\vec{R}_m(t)$ resulting from the entangled initial state in the polaron frame has three contributions: 1) the $R^{(e)}_{mi}$ terms proportional to $\alpha^e_m$,
\begin{eqnarray}\label{Re}
R^{(e)}_{mx}&=&G^{1+}_{mz}\alpha^e_{m},~R^{(e)}_{my}=G^{2-}_{mz}\alpha^e_{m},\nonumber\\
R^{(e)}_{mz}&=&-(G^{1+}_{mx}+G^{2-}_{my})\alpha^e_{m};
\end{eqnarray}
and 2) the conventional first and second order inhomogeneous terms $R^{(1)}_{mi}$ and $R^{(2)}_{mi}$, which are related to an auxiliary function $d_m(t)$ given by
\begin{eqnarray}\label{dm}
d_m(t)=\exp[{\sum_k i{\omega^{-2}_k}{\xi_k \sin\omega_kt(2m\eta_k-\xi_k)}}]-1.
\end{eqnarray}
The explicit expressions for these two contributions are listed in Appendix A. We have mentioned that the dynamics of the TLS is not affected by the boson bath if the T-B coupling vanishes ($\xi_k=0$), which can be readily seen by setting $\phi(t)=0$ and $d_m(t)=0$.

\subsection{Polarization dynamics of the TLS}

In this section, we focus on the polarization dynamics of the TLS. We start with the steady state solutions, i.e., the TLS polarization at long times, as it evolves from an initial down state. As in the spin-boson model, coupling with the boson bath plays a key role in driving the TLS into its steady states. For $\xi_k=0$, the dynamics is always coherent, and there is no steady state, as revealed by the singularity of the Bloch matrix $M(t)$ [by setting $G^{\xi\pm}_{mi}=0$ in Eq.~(\ref{xyz})]. In fact, Eq.~(\ref{Bloch}) gives the probability of finding the TLS in its up state \cite{PRL2012}
\begin{eqnarray}\label{xi0}
P_1(t)&=&\frac {1+\langle\sigma_z\rangle(t)}{2} \nonumber\\
&=&\frac{1}{Z_{\rm S}}\sum_{lm}{J^2e^{-\beta\alpha m} \nu\left(l,\frac{N}{2}\right)}\frac{\sin^2 \omega_m t} {\omega_m^2},
\end{eqnarray}
where $\omega_m = [J^2+(\frac{\varepsilon}{2}+\gamma m)^2]^{-1/2}$, and the summation is over a series of nonnegative oscillating functions.
In this case, it was shown in Ref.~\cite{PRL2012} that the maximal amplitude of the oscillating transition rate can reach its maximum for some finite optimal $\gamma$ on proper times scales. Once the T-B coupling is introduced, steady states can be built up. In general, for large enough $t$, the summation in the oscillating exponential of $d_m(t)$ will average to zero, hence both the first and second order inhomogeneous terms vanish (see Appendix A), and only the first term $R^{(e)}_{mi}$ contributes to the inhomogeneous terms. However, the homogenous relaxation functions $G^{\xi\pm}_{mi}(t)$ will be finite even in the long-time limit. Thus, the $R^{(e)}_{mi}$ part is responsible for the steady state, with the solution
\begin{eqnarray}\label{ss}
\vec{\alpha}_m(\infty)=-[M(\infty)]^{-1}\vec{R}^{(e)}_m(\infty).
\end{eqnarray}
For the separable initial state given by Eq.~(\ref{IS}), it can be easily shown that $\alpha^e_m(0)=e^{-\beta\alpha m}/Z_{\rm S}$ holds for any initial TLS density matrix $\rho_{\rm T}$, i.e., the steady state of the TLS is independent of its initial state.
\begin{figure}
\begin{center}
\includegraphics[scale=0.68, angle=0]{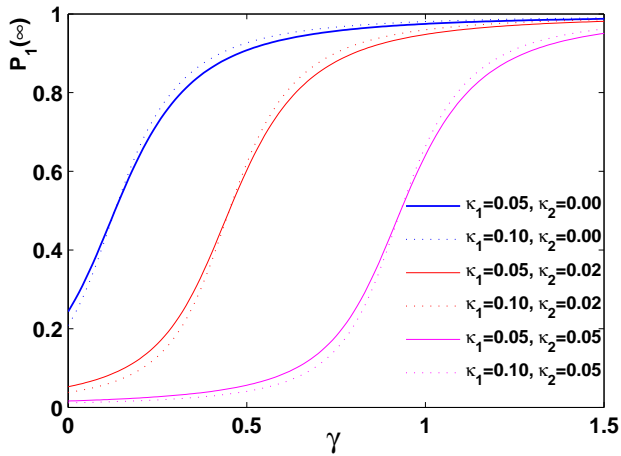}
\caption{Steady-state probability of finding the TLS in its up state as a function of the T-S coupling strength $\gamma$
after prepared in the down state initially. Parameters used are: $\varepsilon=J=\alpha=1,~T=0.5,~N=10$, and $\omega_c=2$.}
\end{center}
\end{figure}

Fig.~2 displays the steady state value of the probability $P_1(t)$ as a function of the T-S coupling strength $\gamma$.
It can be seen that for all cases considered, the steady-state occupation probability $P_1(\infty)$ of the upper level is an increasing function of $\gamma$. This is due to the dominance of the T-S coupling at large $\gamma$. Thermodynamic considerations leave larger Boltzmann weights to spin-bath states of negative $m$'s, driving  $\langle \sigma_z \rangle$ toward unity as $\gamma$ increases. The steady-state probability is robust against variations of the T-B coupling. However, same is not true for variations in the S-B coupling $\kappa_2$. To achieve a given value of $P(\infty)$, a larger $\gamma$ is needed for a larger $\kappa_2$. In this sense, the correlation between the two baths plays a destructive role in the TLS flipping. This can be understood from considering the renormalized T-S coupling $\tilde{\gamma}=\gamma-2\kappa_2\omega^3_c$: a larger $\gamma$ is needed to offset the effect of renormalization due to the S-B coupling.

\begin{figure}
\begin{center}
\includegraphics[scale=0.38, angle=0]{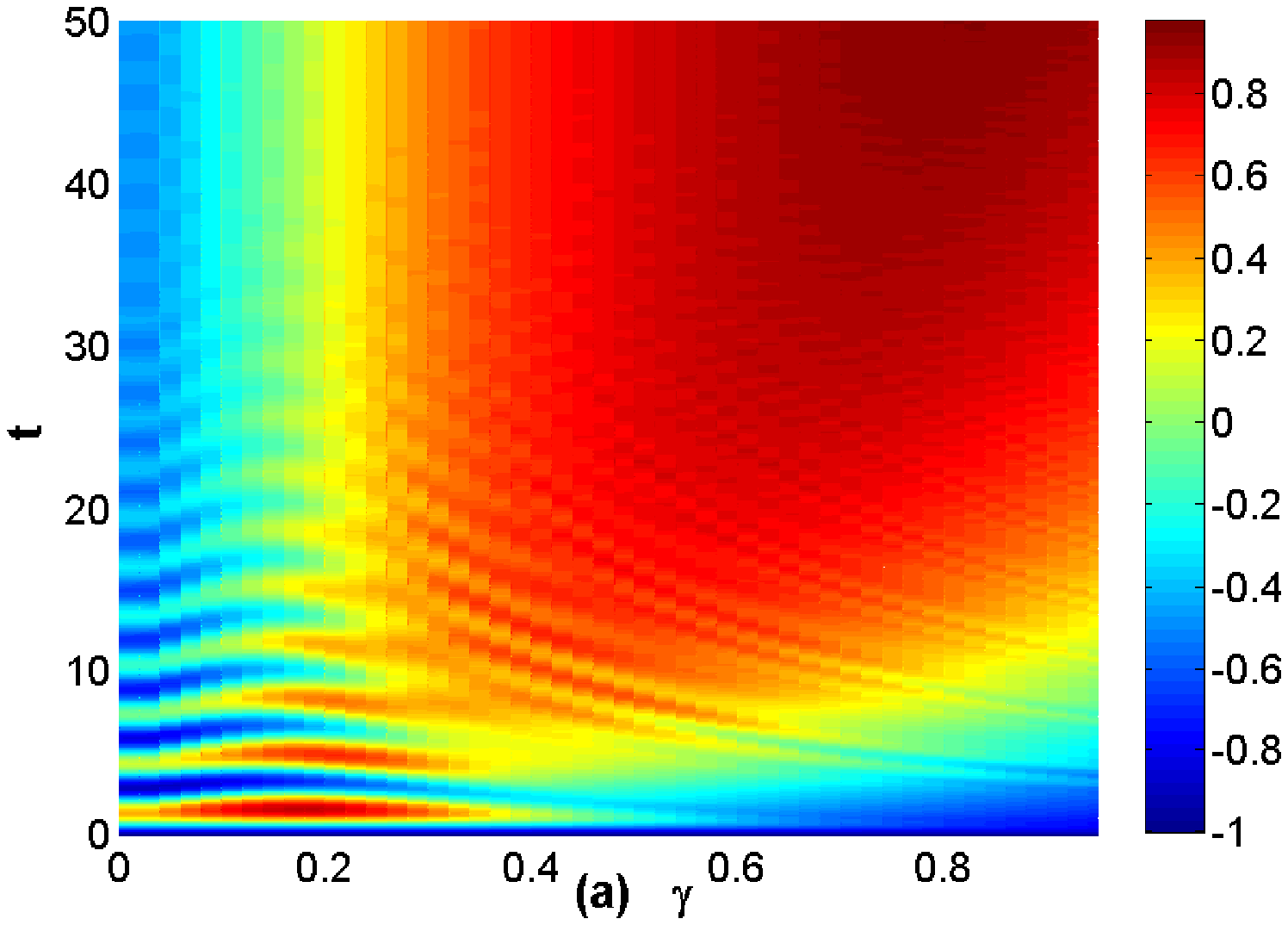}
\includegraphics[scale=0.38, angle=0]{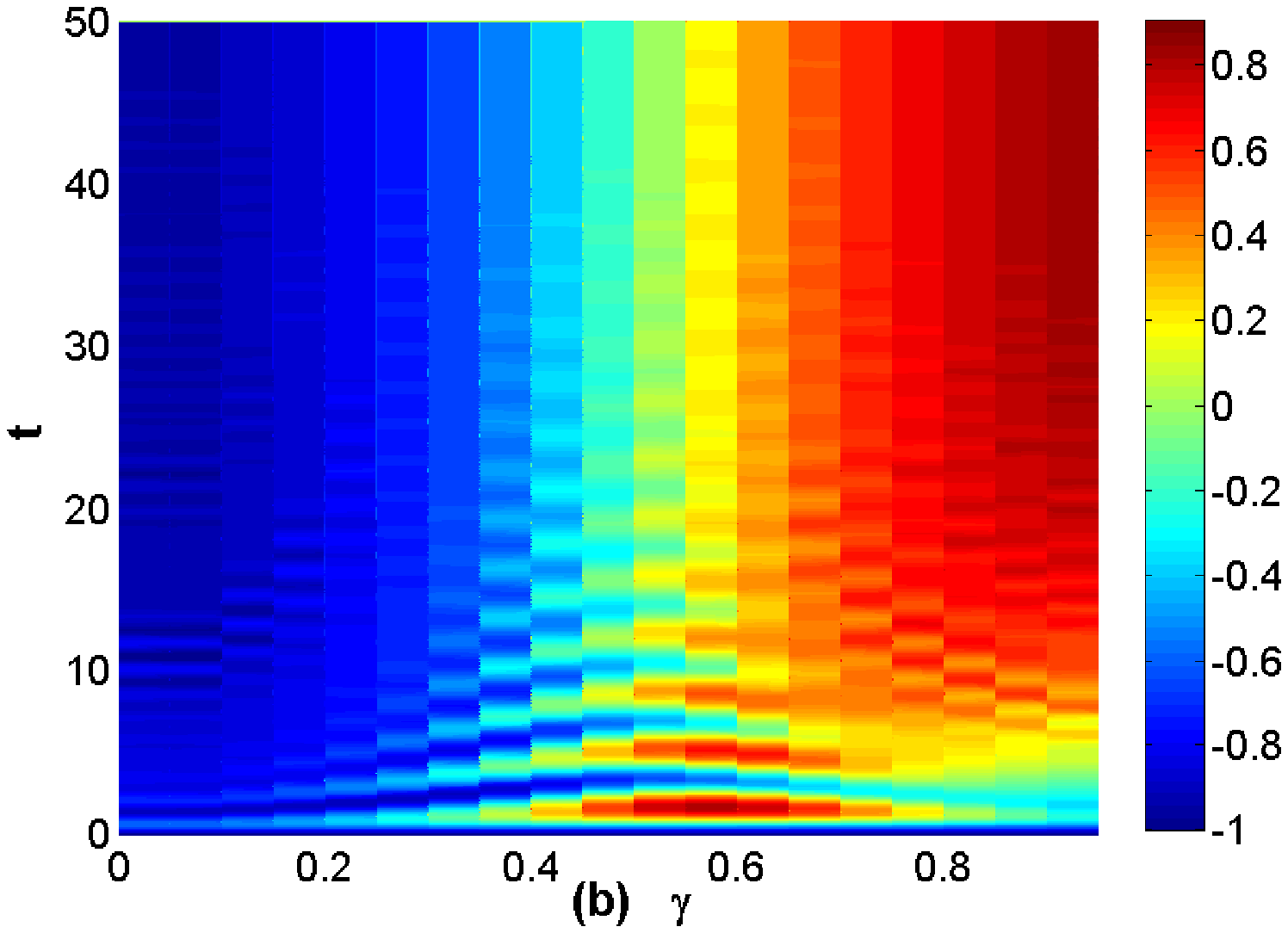}
\caption{the TLS polarization $\langle\sigma_z\rangle(t)$ as a function of time $t$ and the T-S coupling strength $\gamma$ for two values of $\kappa_2$: $\kappa_2=0$ (upper) and $\kappa_2=0.02$ (lower). $\kappa_1=0.05$, and other parameters are the same as those in Fig.~2.}
\end{center}
\end{figure}

In most situations, especially under the Markovian approximation, a small system interacts with a sufficiently large thermal bath, before reaching the eventual thermal equilibrium state with the bath sharing the same temperature \cite{thermal}. However, this may not be the case for an Ohmic spin-boson model with weak system-bath coupling, for which it is found that the spin does not settle to the Gibbs distribution of the uncoupled system if only the first order Born approximation is made \cite{PRB2005}. Instead, the steady state is consistent with a Gibbs distribution of the system-bath combination.
For the TLS described by $H_{\rm T}$ in the thermal equilibrium with inverse temperature $\beta$, the probability in its up state is found to be
\begin{eqnarray}\label{Peq}
P_1^{\rm eq}&=&\frac{1}{2}\left(1+\frac{\rm{Tr}_{\rm T}(e^{-\beta H_T}\sigma_z)}{\rm{Tr}_{\rm T}(e^{-\beta H_T})}\right)\nonumber\\
&=&\frac{1}{2}\left[1- \frac{1}{\sqrt{1+(2J/\varepsilon)^2}}\tanh\beta\sqrt{J^2+\left(\frac{\varepsilon}{2}\right)^2}\right].
\end{eqnarray}
It will be interesting to compare the above expression with the case of $\gamma=0$, where the TLS is only coupled to the boson bath. For parameters used in Fig.~2, we have $P_1^{\rm eq}=0.2814$, which is very close to the steady state value for vanishing S-B coupling ($\kappa_2=0$, see, e.g., the solid blue line in Fig.~2). In this case our model is reduced to the traditional spin-boson model. The slight variance of $P_1 (\infty)$ with $\kappa_1$ may be due to non-Markovian effects related to the TCL master equation approach.
However, even though there are no direct interactions between the TLS and the spin bath, $P_1(\infty)$ deviates from the equilibrium value for finite $\kappa_2$.

We can view the TLS and the spin bath as two sub-systems that couple to a common boson bath. It is intriguing that the spin bath with only 10 spins in our numerics and no direct coupling with the TLS displays noticeable influences over the thermalization of the TLS. Qualitatively, the interbath S-B coupling will be dominant for large $\kappa_2$. In contrast to the T-S coupling term, $\gamma L_z\sigma_z$, which drives the polarization of the TLS to be anti-parallel with respect to that of the spin bath, the ``effective magnetic field," $b_k+b^\dag_k$, will align $\sigma_z$ and $L_z$ into a parallel formation. Using the energy-based argument, it can be concluded that a larger S-B coupling strength tend to yield a smaller $P_1(\infty)$.
In general, dynamics induced by a spin bath is intrinsically non-Markovian \cite{Breuer-2004}, and the spin bath may more likely drive the TLS out of thermalization in a steady state. In fact, deviation of the steady state from the the Gibbs equilibrium is reported recently in Ref.~\cite{Zhou} for a spin-star system. Furthermore, due to the non-Markovian nature of the spin bath, the steady state is more dependent on the initial state of the TLS.

To study the time dependence of the population difference $\langle\sigma_z\rangle(t)$, we have solved Eq. (\ref{Bloch}) numerically. In the upper panel of Fig.~3, $\langle\sigma_z\rangle(t)$ as a function of time $t$ and the T-S coupling constant $\gamma$ is presented for $\kappa_1=0.05$ and $\kappa_2=0$. Similar results are displayed in the lower panel of Fig.~3 for $\kappa_1=0.05$ and $\kappa_2=0.02$.
It is revealed that an optimized value of $\langle\sigma_z\rangle(t)$ can be found in both cases: for $\kappa_2=0$, it occurs at $\gamma\approx 0.2$; while for $\kappa_2=0.02$, at $\gamma\approx0.6$. This phenomena is also observed when the T-B coupling is absent\cite{PRL2012}. The shift of optimal T-S coupling strength for different bath correlations again results from the renormalized $\tilde{\gamma}$.
Interesting dynamics emerges in the large-$\gamma$ regime, where the steady-state population transfer increases monotonically with $\gamma$, and for a given $\gamma$,
the TLS flips
unidirectionally barring minor population oscillations of minute amplitudes, as shown in Fig.~3.
Overall, the cooperative interplay of the T-B and  T-S coupling facilitates efficient, unidirectional energy transport.

\section{Macroscopic quantum-superposition states of the spin bath}

So far the reduced dynamics of the TLS has been our focus. In this picture, the spin bath is viewed as an alternative bath which induces decoherence and dissipation along with the conventional boson bath. On the other hand, the spin bath in our system can also model an atomic ensemble that is coupled to the boson bath.
For example, the dynamics of the two-mode Bose-Hubbard model can be obtained by mapping the bosonic system onto one of the spins \cite{PRA2003,PRL2008,PRA2010}. Using an exactly solvable model, it was demonstrated that the Hamiltonian $H_2$ can drive an uncorrelated multi-spin system into a macroscopic quantum-superposition (MQS) state  with high fidelity \cite{kurizki}.
The concept of MQS \cite{PRL1986,PRA2003,kurizki} in multi-(pseudo)spin or atomic systems is of great interest. For a system made of $N$ non-interacting spins-1/2 described by the collective operator $\mathbf{L}=\sum^N_{j=1}\vec{\sigma}_j/2$, take a spin coherent state as its initial state such that $\rho_{\rm sc}(\theta,\phi,\gamma)=|\hat{\Omega}\rangle\langle\hat{\Omega}|$ with $|\hat{\Omega}\rangle=e^{-iL_z\phi}e^{-iL_y\theta}e^{-iL_z\gamma}|{N}/{2},{N}/{2}\rangle$, which is peaked along the direction $\hat{\Omega}=(\sin\theta\cos\phi,\sin\theta\sin\phi,\cos\theta)$. By setting the gauge angle $\gamma$ to zero, the spin coherent state can be written in the $|j,m\rangle$ basis as\cite{MQM}
\begin{eqnarray}\label{SC}
|\hat{\Omega}\rangle=\sqrt{N!}\sum^{N/2}_{m=-N/2}\frac{u^{\frac{N}{2}+m}v^{\frac{N}{2}-m}}{\sqrt{(\frac{N}{2}-m)!(\frac{N}{2}+m)!}}|\frac{N}{2},m\rangle,
\end{eqnarray}
where $u=\cos\frac{\theta}{2}e^{-i {\phi}/{2}}, v=\sin\frac{\theta}{2}e^{i {\phi}/{2}}$. Specifically, states $|\pm\hat{x}\rangle$ with all spins in the spin bath pointing to the $\pm\hat{x}$ axis can be obtained by choosing $(\theta,\phi)=(\frac{\pi}{2},\frac{\pi}{2}\mp\frac{\pi}{2})$:
\begin{eqnarray}
|\pm \hat{x}\rangle=\sum^{N/2}_{m=-N/2}(\pm1)^mq_m|\frac {N}{2},m\rangle
\end{eqnarray} where
$q_m=2^{- {N}/{2}}\sqrt{C^{ {N}/{2}+m}_N}$.
Within a model described by $H_2$, Ref.~\cite{kurizki} examined the reduced dynamics of the spin bath evolving from an initial state
\begin{eqnarray}\label{rhoL0}
\Theta^{(\frac{N}{2})}_{\rm S}(0)=|+\hat{x}\rangle\langle+\hat{x}|=\sum_{mn}q_mq_n |\frac{N}{2},m\rangle\langle\frac{N}{2},n|,
\end{eqnarray}
where $\Theta^{(l)}_{\rm S}(t)$ denotes the reduced density matrix of the spin bath in the $l$-subspace. It is found that, when the decoherence rate caused by the boson bath is negligible, $\Theta^{(\frac{N}{2})}_S(t)$ can be approximated by \cite{kurizki}
\begin{eqnarray}\label{rhoLt}
\Theta^{(\frac{N}{2})}_{\rm S}(t)&\approx& \sum_{mn}q_mq_n e^{-itf(t)(m^2-n^2)}|\frac{N}{2},m\rangle\langle\frac{N}{2},n|,\nonumber\\
f(t)&=&\sum_k\frac{\eta^2_k}{\omega_k}(1-\frac{\sin\omega_kt}{\omega_kt}).
\end{eqnarray}
Comparing Eq.(\ref{rhoLt}) with Eq.(\ref{rhoL0}), it is found that the system returns to its initial state when $t= {2\pi}/{f(t)}$, that is, the state $\Theta^{(\frac{N}{2})}_S(t)$ is periodic with a period of $  {2\pi}/{f(t)}$. When $t= {\pi}/{f(t)}$, we have $e^{-itf(t)(m^2-n^2)}=(-1)^{m^2-n^2}=(-1)^{m+n}$, then the state evolves into the state $|-\hat{x}\rangle$. If we defined $\tau_{\rm MQS} \equiv {\pi}/{2f(t)}$, of great interest is what happens at $t=\tau_{\rm MQS}$. It can be easily checked that the state evolves into an entangled MQS state \cite{PRL1986}
\begin{eqnarray}\label{psiMQS}
|\psi_{\rm MQS}\rangle &=&\frac{1}{\sqrt{2}}(|+\hat{x}\rangle+i|-\hat{x}\rangle),\\
\rho_{\rm MQS}&\equiv&|\psi_{\rm MQS}\rangle\langle\psi_{\rm MQS}|.
\end{eqnarray}
Apparently, a perfect MQS state has matrix elements
$|[\rho_{\rm MQS}]_{\pm\pm}|=|\langle\pm\hat{x}|\rho_{\rm MQS}|\pm\hat{x}\rangle|=\frac{1}{2}$. When the decoherence rate is included, the MQS can still be achieved with high probability at sufficiently low temperatures \cite{kurizki}.

Our goal here is to study the effects of the TLS on the MQS generation. Along this line, TLS-induced correlations and entanglements have been studied previously for a bath of spins that is coupled to the TLS via XX-type coupling in a spin-star configuration \cite{Bose-2002}.
In this work, the reduced dynamics of the spin bath is probed by tracing out the degrees of freedom in the TLS and the boson bath.
We are mainly interested in the matrix elements
\begin{eqnarray}\label{thetapp}
\Theta_{\pm\pm}(t)&=&\langle\pm \hat{x}|\Theta^{(\frac{N}{2})}_{\rm S}(t)|\pm \hat{x}\rangle\nonumber\\
&=&\sum_{mn}(\pm1)^m(\pm1)^nq_mq_n[\Theta^{(\frac{N}{2})}_{\rm S}(t)]_{mn}.
\end{eqnarray}
Finite values of the above matrix elements signify the presence of an MQS state with high fidelity.
In Eq.~(\ref{thetapp}), the matrix elements of the reduced density matrix $\Theta^{(\frac{N}{2})}_{\rm S}(t)$ can be evaluated as
\begin{eqnarray}\label{Thetamn}
&&[\Theta^{(\frac{N}{2})}_{\rm S}]_{mn}=\langle \frac{N}{2},m|\Theta^{(\frac{N}{2})}_{\rm S}|\frac{N}{2},n\rangle={\rm Tr}_{\rm T+B}\nonumber\\
&&\{\langle \frac{N}{2},m|\tilde{\rho}_I(t)|\frac{N}{2},n\rangle e^{i(H_n-H_m)t}e^{(n-m)B_1(t)}\},
\end{eqnarray}
where the TLS operator
\begin{eqnarray}\label{Hm}
H_m=(\alpha m-\eta m^2)+{\epsilon_m}\frac{\tau^z_m}{2}
\end{eqnarray}
satisfies $\tilde{H}_0|lm\rangle=|lm\rangle H_m$.
We leave the details of evaluating $[\Theta^{(\frac{N}{2})}_{\rm S}]_{mn}$ using the TCL master equation to Appendix B.

 Following Ref.~\cite{kurizki}, we take our initial state of the whole system as a separable state
\begin{eqnarray}\label{rho00}
\rho(0)=|\bar{1}\rangle\langle\bar{1}|\otimes|+\hat{x}\rangle\langle+\hat{x}|\otimes\rho_{\rm B}.
\end{eqnarray}
or, after transforming into the polaron frame,
\begin{eqnarray}\label{rho00P}
\tilde{\rho}(0)&=&\sum_{mn}q_mq_n|\bar{1} m\rangle\langle\bar{1 }n|\tilde{\rho}_{\rm B}^{mn},\nonumber\\
\tilde{\rho}^{mn}_{\rm B}&=&e^{mB_1-\frac{1}{2}B_2}\rho_{\rm B}e^{-nB_1+\frac{1}{2}B_2}.
\end{eqnarray}

\begin{figure}[tb]
\begin{center}
\includegraphics[scale=0.45, angle=0]{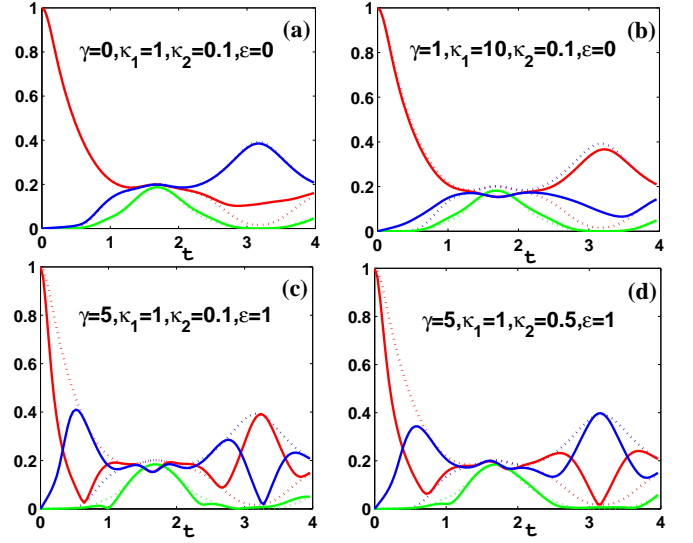}
\caption{Time-dependent magnitudes of matrix elements $|\Theta_{++}|$ (red), $|\Theta_{+-}|=|\Theta_{-+}|$ (green), and $|\Theta_{--}|$ (blue). Parameters chosen: $N=10, J=0.1,~\alpha=0,~\omega_c=1,~\beta=100$ and $\kappa_3=0.5$. The dotted lines are calculated in the absence of the TLS\cite{kurizki} and $\tau_{\rm MQS}=1.685$. }
\end{center}
\end{figure}

In Fig.~4, we display for four sets of parameters the time-dependent magnitudes of the four matrix elements $|\Theta_{++}|$,$|\Theta_{+-}|=|\Theta_{-+}|$ and $|\Theta_{--}|$ in Eq.~(\ref{thetapp}). Simultaneous deviation of those matrix elements from zero signifies the formation of an MQS state with high fidelity, as mentioned earlier.
In the absence of the TLS, the MQS state can be reached at $\tau_{\rm MQS}\approx 1.685$ with a high probability \cite{kurizki}, where the parameters chosen are $J=0.1,~\alpha=0,~\omega_c=1,~\beta=100$ and $\kappa_3=0.5$, in order to keep the temperature sufficiently low for superposition generation, namely. For comparison, results in the absence of the TLS are shown as dotted lines in Fig.~4.
It is clearly seen that the initial state $|+\hat{x}\rangle$ evolves into an MQS state, as given by Eq.~(\ref{psiMQS}), with a high probability as indicated by $|\Theta_{++}(\tau_{\rm MQS})|\approx|\Theta_{+-}(\tau_{\rm MQS})|\approx|\Theta_{-+}(\tau_{\rm MQS})|\approx|\Theta_{--}(\tau_{\rm MQS})|\approx0.2$, a much reduced value when compared to 0.5 for a perfect MQS state. The considerable drop of the matrix elements from 0.5 to 0.2 is attributed to the boson-bath decoherence, and at $t\approx2\tau_{\rm MQS}$, the state evolves to $|-\hat{x}\rangle$ with a high probability as indicated by $|\Theta_{--}(2\tau_{\rm MQS})|\approx0.4$, and $|\Theta_{++}(2\tau_{\rm MQS})|\approx|\Theta_{+-}(2\tau_{\rm MQS})|\approx|\Theta_{-+}(2\tau_{\rm MQS})|\approx0$.

Unlike in the study of the reduced dynamics of the TLS, where the bath-bath correlation is reflected from the hybridization of the T-B and S-B coupling through the spectral function $J_{\rm TS}(\omega)$, direct bath-bath interactions play a role via $J_{\rm SS}(\omega)$ in the present case. The nonlinear term, $-L^2_z\eta$, in Eq.~(\ref{eq5}) is the driving force for the formation of an MQS state \cite{kurizki}, while such a term has no effect on the reduced dynamics of the TLS.
Comparison between Fig.~4(a) and Fig.~4(b) indicates that the T-B coupling has only a minor impact on the dynamics of the matrix elements before $t =\tau_{\rm MQS}$, except for an inversion from state $|-\hat{x}\rangle$ to $|+\hat{x}\rangle$ at $t\approx2\tau_{\rm MQS}$ as the T-B coupling is increased and the system goes from the weak to strong coupling regime. A similar inversion can be found in Fig.~4(c) when $\gamma$ is increased to 5. Such inversions roughly reduce the quasi-period of the evolution from $4\tau_{\rm MQS}$ to $2\tau_{\rm MQS}$ without changing the onset time for the MQS state.

However, finite T-S coupling $\gamma$ or energy difference $\varepsilon$ have considerable effects on the diagonal elements $|\Theta_{++}|$ and $|\Theta_{--}|$.
As shown in Figs.~4(c) and (d), the state evolves approximately into a fully mixed state with a density matrix in the $\{|+\hat{x}\rangle,|-\hat{x}\rangle\}$ basis given by
\begin{eqnarray}
\rho_{\rm mix}=\left(
  \begin{array}{cc}
  \frac{1}{2} & 0 \\
  0 & \frac{1}{2} \\
   \end{array}
   \right).
    \end{eqnarray}
When the solid blue and red lines cross Figs.~4(c) and (d), one has approximately that $|\Theta_{++}|=|\Theta_{--}|\approx0.2$ and $|\Theta_{+-}|=|\Theta_{-+}|\approx0$. Such mixed states emerge even before the inversion from $|+\hat{x}\rangle$ to $|-\hat{x}\rangle$ and the MQS generation (i.e, $t < \tau_{\rm MQS}$). Take Fig.~4(c) as an example, the state evolution cycle is found to follow roughly:
\begin{eqnarray}
&&~~~~|+\hat{x}\rangle\langle+\hat{x}| (t=0)\nonumber\\
&&\to\rho_{\rm mixed} (t\approx\tau_{\rm MQS}/4)\nonumber\\
&&\to|+\hat{x}\rangle\langle+\hat{x}|(t\approx\tau_{\rm MQS}/2)\nonumber\\
&&\to\rho_{\rm MQS}(t\approx\tau_{\rm MQS})\nonumber\\
&&\to|-\hat{x}\rangle\langle-\hat{x}|(t\approx3\tau_{\rm  MQS}/2)\nonumber\\
&&\to\rho_{\rm mixed}(t\approx7\tau_{\rm  MQS}/4)\nonumber\\
&&\to|+\hat{x}\rangle\langle+\hat{x}|(t\approx2\tau_{\rm MQS}).\nonumber
\end{eqnarray}
Despite the rich dynamical behavior of the diagonal elements $|\Theta_{++}|$ and $|\Theta_{--}|$, the off-diagonal elements, $|\Theta_{+-}|=|\Theta_{-+}|$ (the green lines), are insensitive to the changes in control parameters, and to the removal or addition of the TLS, which pins $\tau_{\rm MQS}$ to roughly 1.7.
It is thus concluded that the generation of macroscopic quantum superposition states is robust under various changes of parameter sets including the removal of the TLS. Note that, as can be expected, the additional effect imposed by the TLS is of the order of $O(1/N)$ and will become negligible as the number of spins $N$ in the spin bath goes to infinite.

\section{Conclusions}

An extension to the spin-boson model has been proposed by including an additional spin bath in a spin-star configuration in an effort to study: 1) the polarization dynamics of the TLS under the influence of the two dissimilar baths; and 2) the macroscopic quantum superposition state generation in the spin bath.
By incorporating two limiting cases in which the dynamics can be solved exactly, effects of the two baths on the TLS have been investigated first.
Steady states can be reached in the presence of a boson bath with a continuous spectral density.
It is found that the steady flipping of the TLS is aided by a finite TLS-spin-bath coupling strength.
This can be understood from thermodynamic considerations as spin-bath states with a negative magnetic number $m$ has a large Boltzmann weight favoring a unity value of $\langle\sigma_z\rangle$ with the increasing T-S coupling $\gamma$. Furthermore, the bath-bath correlation $\kappa_2$ plays a destructive role in the TLS flipping.
More interestingly, the spin bath has a sizable effect on TLS thermalization even when there is no direct interaction between the TLS and the spin bath. For the transient process, analogous to the findings in Ref.~\cite{PRL2012}, optimal flipping occurs at an intermediate T-S coupling strength, the value of which is sensitive to the bath-bath correlation.
For large enough T-S coupling, the TLS flips unidirectionally barring minor oscillations of minute amplitudes. Lastly, we have studied the MQS state generation in the spin bath under the influence of the boson bath and the TLS.
Given that MQS can be achieved in the absence of the TLS, the effect of adding the TLS on MQS has been investigated.
In the presence of strong T-B coupling, the quasi-period of evolution can be reduced roughly by half. Rich dynamics can appear as the T-S coupling is increased, such as the emergence of a fully mixed state of $|+\hat{x}\rangle$ and $|+\hat{x}\rangle$ before $\tau_{\rm MQS}$ is reached, a phenomenon that is absent if the spin bath is only coupled to the boson bath (without the TLS). It is also found that the generation of MQS states is robust against parameter variance including the removal of the TLS due to the insensitivity of the off-diagonal elements of the reduce density matrix.

\section*{Acknowledgements}

The authors gratefully acknowledge support from the the Singapore National Research
Foundation through the Competitive Research Programme (CRP) under
Project No.~NRF-CRP5-2009-04.

\appendix
\section{Inhomogeneous terms in Eq.(\ref{xyz})}
The first order inhomogeneous terms $R^{(1)}_{mi}$ in Eq.(\ref{xyz}) are given by
\begin{eqnarray}
R^{(1)}_{mx}&=&-\Theta d^{-}_m\alpha^e_{m}(S^2_m\cos\epsilon_mt+C^2_m),\nonumber\\
R^{(1)}_{my}&=&\Theta d^{+}_m\alpha^e_{m}(S^2_m\cos\epsilon_mt+C^2_m),\nonumber\\
R^{(1)}_{mz}&=&\Theta S_m\alpha^e_m[\sin\epsilon_mt d^+_m-C_m(\cos\epsilon_mt-1)d^-_m].\nonumber\\
\end{eqnarray}
where $d^+_m=d_m+d^*_m,~d^-_m=i(d_m-d^*_m)$.
\par The second order inhomogeneous terms $R^{(2)}_{mi}$ in Eq.(\ref{xyz}) are given by
\begin{eqnarray}
R^{(2)}_{mx}&=&\alpha^e_m[G^{1-}_{Qmx}S_mC_m\sin\epsilon_mt-2S_mG^{1+}_{Qmx}\nonumber\\
&&+G^{1-}_{Qmy}S_m C_m(1-\cos\epsilon_mt)+2G^{1+}_{Qmz}C_m\nonumber\\
&&+G^{1+}_{Qmz}S^2_m\sin\epsilon_mt],\nonumber\\
R^{(2)}_{my}&=&\alpha^e_m[-G^{2+}_{Qmx}S_mC_m\sin\epsilon_mt-2S_mG^{2-}_{Qmx}\nonumber\\
&&-G^{2+}_{Qmy}S_m C_m(1-\cos\epsilon_mt)+2G^{2-}_{Qmz}C_m\nonumber\\
&&-G^{2+}_{Qmz}S^2_m\sin\epsilon_mt],\nonumber\\
R^{(2)}_{mz}&=&\alpha^e_m[-2C_mG^{1+}_{Qmx}-S^2_m\sin\epsilon_mt G^{1-}_{Qmx}\nonumber\\
&&+C_mG^{2+}_{Qmx}+(C^2_m+S^2_m\cos\epsilon_mt)G^{1-}_{Qmy}\nonumber\\
&&-2G^{2-}_{Qmy}-2S_mG^{1+}_{Qmz}+S_mC_m\sin\epsilon_mtG^{1-}_{Qmz}\nonumber\\
&&+S_m\cos\epsilon_mt G^{2+}_{Qmz}],
\end{eqnarray}
where the inhomogeneous rates $G^{\xi+}_{Qmi}=J^2(\gamma^\xi_{Qmi}+\gamma^{\xi*}_{Qmi})$ and $G^{\xi-}_{Qmi}=iJ^2(\gamma^\xi_{Qmi}-\gamma^{\xi*}_{Qmi})$ are combinations of
\begin{eqnarray}\label{iHcf}
&&\gamma^{1,2}_{Qmi}=\nonumber\\
&&\Theta^2\int^t_0ds\{e^{-\phi(t-s)}[K^i_m(s-t)d_m(t)d_m(s)\mp c.c.]\nonumber\\
&&+(e^{-\phi(t-s)}-1)[K^i_m(s-t)(d_m(t)+d_m(s))\mp c.c.]\}\nonumber\\
&&+\Theta^2\int^t_0ds\{e^{\phi(t-s)}[K^{i*}_m(s-t)d_m(t)d_m^*(s)\mp c.c.]\nonumber\\
&&+(e^{\phi(t-s)}-1)[K^{i*}_m(s-t)(d_m(t)+d_m^*(s))\mp c.c.]\}.\nonumber\\
\end{eqnarray}
Note that, unlike $R^{(e)}_{mi}$, these two terms are directly related to $d_m(t)$. Hence the $R^{(1)}_{mi}$ and $R^{(2)}_{mi}$ terms will vanish once $d_m(t)$ vanishes, which is the case in the long-time limit.

\section{Reduced dynamics of the spin bath}
In this Appendix we describe the calculations of the relevant and irrelevant parts of the matrix elements $[\Theta^{(\frac{N}{2})}]_{mn}$, from which $\Theta_{\pm\pm}(t)$ can be obtained from Eq.(\ref{thetapp}). Similar to the bath correlation function $\phi(t)=\phi_1(t)-i\phi_2(t)$ in Eq. (\ref{CF}), the following correlation functions
\begin{eqnarray}
\psi(t)&=&\psi_1(t)-i\psi_2(t)\nonumber\\
&=&\sum_k\left(\frac{\eta_k}{\omega_k}\right)^2\left(\cos\omega_kt\coth\frac{\beta\omega_k}{2}-i\sin\omega_kt\right),
\end{eqnarray}
and
\begin{eqnarray}
\psi_{mn}(t)=(m-n)\sum_k\frac{\eta_k\xi_k}{\omega^2_k}\coth\frac{\beta\omega_k}{2}\cos\omega_kt,
\end{eqnarray}
are found to be useful. For the cubic spectral density used in this work, we have
\begin{eqnarray}
\psi_1(t)&=&\frac{\kappa_3}{\kappa_1}\phi_1(t),\nonumber\\
\psi_{m-n}(t)&=&(m-n)\frac{\kappa_2}{\kappa_1}\phi_1(t).
\end{eqnarray}
By applying the identity $\mathcal{P}+\mathcal{Q}=1$ to $\tilde{\rho}_I(t)$ in Eq.(\ref{Thetamn}), we can divide $[\Theta^{(l)}_{\rm S}]_{mn}$ into the relevant part and the irrelevant part.
It is known that the irrelevant part $\mathcal{Q}\tilde{\rho}_{\rm I}(t)$ can be determined from $\mathcal{Q}\tilde{\rho}_{\rm I}(0)$ and $\mathcal{P}\tilde{\rho}_{\rm I}(t)$ as $\mathcal{Q}\tilde{\rho}_{\rm I}(t)=[1+A(t)]\mathcal{Q}\tilde{\rho}_{\rm I}(0)+B(t)\mathcal{P}\tilde{\rho}_{\rm I}(t)$, with $A(t)$ and $B(t)$ operators of the first order in $\tilde{H}_{1,I}$ \cite{Breuer}.  In practice, an exact calculation of the irrelevant part can be very involved. Here we will use the zeroth order approximation $\mathcal{Q}\tilde{\rho}_{\rm I}(t)\approx\mathcal{Q}\tilde{\rho}_{\rm I}(0)$\cite{Nazir2011}, so that
\begin{eqnarray}
&&[\Theta^{(\frac{N}{2})}_{\rm S}]_{mn}=[\Theta^{(\frac {N}{2})}_{\rm S}]^P_{mn}+[\Theta^{(\frac{N}{2})}_{\rm S}]^Q_{mn},
\end{eqnarray}
with
\begin{eqnarray}\label{ThetaP}
[\Theta^{(\frac{N}{2})}_{\rm S}]^P_{mn}&=&{\rm Tr}_{\rm T}\{ \langle  m| \tilde{\sigma}_{\rm I}(t)| n\rangle e^{i(H_n-H_m)t}\}\nonumber\\
 &&{\rm Tr}_{\rm B}\{\rho_{\rm B}e^{(n-m)B_1(t)}\},
\end{eqnarray}
 and
\begin{eqnarray}\label{ThetaQ}
&&[\Theta^{(\frac{N}{2})}_{\rm S}]^Q_{mn}\approx\nonumber\\
&&{\rm Tr}_{\rm T+B}\{ \langle  m| \mathcal{ Q} \tilde{\rho}_{\rm I}(0)| n\rangle e^{i(H_n-H_m)t} e^{(n-m)B_1(t)}\}.
\end{eqnarray}
The irrelevant part is evaluated directly for a given initial state,
while the relevant part can be obtained by using the TCL master equation on matrix elements of  $\tilde{\sigma}_{\rm I} (t)$
\begin{eqnarray}\label{hsm}
h_{sm,s'm'}(t)\equiv \langle sm|\tilde{\sigma}_{\rm I}(t)|s'm'\rangle,
\end{eqnarray}
where we have defined the basis $|sm\rangle \equiv |s\rangle |{N}/{2},m\rangle $, with $s=1,\bar{1}$.

\subsection{Calculation of $[\Theta^{(\frac{N}{2})}]^P_{mn}$ in Eq.(\ref{ThetaP})}

In order to evaluate the trace over the  qubit in Eq. (\ref{ThetaP}), we diagonalize $H_n-H_m$ in the $\sigma$-basis:
\begin{eqnarray}
H_n-H_m=\alpha(n-m)-\eta(n^2-m^2)+\frac{1}{2}E_{nm}\tau^z_{nm},
\end{eqnarray}
where $\tau^z_{nm}=|+\rangle_{nm}~_{nm}\langle+|-|-\rangle_{nm}~_{nm}\langle-|$, with the two eigenstates
\begin{eqnarray}
|+\rangle_{nm}&=&\cos\frac{\theta_{nm}}{2}|1\rangle+\sin\frac{\theta_{nm}}{2}|\bar{1}\rangle,\nonumber\\
|-\rangle_{nm}&=&-\sin\frac{\theta_{nm}}{2}|1\rangle+\cos\frac{\theta_{nm}}{2}|\bar{1}\rangle.
\end{eqnarray}
where $\tan\theta_{nm}=(\epsilon_n S_n -\epsilon_m S_m)/(\epsilon_nC_n-\epsilon_mC_m)$.
\par The corresponding eigen-energies read
\begin{eqnarray}
E_{nm,\pm}=\alpha(n-m)-\eta(n^2-m^2)\pm\frac{1}{2}E_{nm},
\end{eqnarray}
with $E_{nm}=\sqrt{(\epsilon_nC_n-\epsilon_mC_m)^2+(\epsilon_n S_n -\epsilon_m S_m)^2}$.
\par By defining $h_{sm,s'm'}=\langle sm|\tilde{\sigma}_{\rm I}(t)|s'm'\rangle$, we obtain
\begin{eqnarray}
&&[\Theta^{(\frac{N}{2})}_{\rm L}]^P_{mn}=e^{-\frac{(n-m)^2}{2}\psi_1(0)}\cdot\nonumber\\
&&[(\cos^2\frac{\theta_{nm}}{2}e^{iE_{nm,+}t}+\sin^2\frac{\theta_{nm}}{2}e^{iE_{nm,-}t})h_{1m,1n}+\nonumber\\
&&\cos \frac{\theta_{nm}}{2}\sin \frac{\theta_{nm}}{2}(e^{iE_{nm,+}t}- e^{iE_{nm,-}t})(h_{1m,\bar{1}n}+h_{\bar{1}m,1n})\nonumber\\
&&+(\sin^2\frac{\theta_{nm}}{2}e^{iE_{nm,+}t}+\cos^2\frac{\theta_{nm}}{2}e^{iE_{nm,-}t})h_{\bar{1}m,\bar{1}n}].
\end{eqnarray}
Thus, we only need to calculate $h_{sm,s'm'}$. Using the TCL master equation, we can construct the equations of motion of $h_{sm,s'm'}$:
\begin{eqnarray}\label{hEOM}
\dot{h}_{sm,s'm'}=[\dot{h}_{sm,s'm'}]_{h}+[\dot{h}_{sm,s'm'}]_{ih1}+[\dot{h}_{sm,s'm'}]_{ih2}
\end{eqnarray}
where $[\dot{h}_{sm,s'm'}]_{h}$, $[\dot{h}_{sm,s'm'}]_{ih1}$, and $[\dot{h}_{sm,s'm'}]_{ih2}$ are the homogeneous term, the first order inhomogeneous term, and the second order inhomogeneous term, respectively:
\begin{eqnarray}
&&[\dot{h}_{sm,s'm'}]_{h}=\nonumber\\
&&-\int^t_0d\tau
{\rm Tr}_{\rm B}\{\langle sm|[\tilde{H}_{1,I}(t),\tilde{H}_{1,I}(\tau)\tilde{\sigma}_{\rm I}(t)\rho_{\rm B}]|s'm'\rangle\nonumber\\
&&+\langle sm|[\tilde{H}_{1,I}(t),\tilde{H}_{1,I}(\tau)\tilde{\sigma}_{\rm I}(t)\rho_{\rm B}]^\dag|s'm'\rangle\}\nonumber\\
&=&2J^2\int^t_0d\tau\{ss'h_{sm,s'm'}\cdot(\Re[\tilde{K}^z_m(\tau)\tilde{K}^z_{m'}(t)]\langle D(t)D(\tau)\rangle\nonumber\\
&&+\Re[\tilde{K}^{z*}_m(\tau)\tilde{K}^z_{m'}(t)]\langle D(t)D^\dag(\tau)\rangle+t\leftrightarrow\tau)\nonumber\\
&+&s'h_{\bar{s}m,s'm'}\cdot\nonumber\\
&&[(\Re[\tilde{K}^z_{m'}(t) \tilde{K}^x_m(\tau)] -is\Re[\tilde{K}^z_{m'}(t)\tilde{K}^y_m(\tau)])\langle D(t)D(\tau)\rangle+\nonumber\\
&&(\Re[\tilde{K}^{z*}_{m'}(t)\tilde{K}^x_m(\tau)] -is\Re[\tilde{K}^{z*}_{m'}(t)\tilde{K}^y_m(\tau)])\langle D^\dag(t)D(\tau)\rangle\nonumber\\
&&+t\leftrightarrow\tau]\nonumber\\
&+&sh_{sm,\bar{s}'m'}\nonumber\\
&&[(\Re[\tilde{K}^z_m(\tau)\tilde{K}^x_{m'}(t)] +is'\Re[\tilde{K}^z_m(\tau)\tilde{K}^y_{m'}(t)])\langle D (t)D (\tau)\rangle+\nonumber\\
&&(\Re[\tilde{K}^z_m(\tau)\tilde{K}^{x*}_{m'}(t)] +is'\Re[\tilde{K}^z_m(\tau)\tilde{K}^{y*}_{m'}(t)])\langle D^\dag (t)D (\tau)\rangle+\nonumber\\
&&t\leftrightarrow\tau]\nonumber\\
&+&  h_{\bar{s}m,\bar{s}'m'}[(\Re[\tilde{K}^x_m(\tau)\tilde{K}^x_{m'}(t)]-is\Re[\tilde{K}^y_m(\tau)\tilde{K}^x_{m'}(t)]\nonumber\\
&&+is'\Re[\tilde{K}^x_m(\tau)\tilde{K}^y_{m'}(t)]+ss'\Re[\tilde{K}^y_m(\tau)\tilde{K}^y_{m'}(t)])\langle D (t)D  (\tau)\rangle\nonumber\\
&&+(\Re[\tilde{K}^x_m(\tau)\tilde{K}^{x*}_{m'}(t)]-is\Re[\tilde{K}^y_m(\tau)\tilde{K}^{x*}_{m'}(t)]\nonumber\\
&&+is'\Re[\tilde{K}^x_m(\tau)\tilde{K}^{y*}_{m'}(t)]+ss'\Re[\tilde{K}^y_m(\tau)\tilde{K}^{y*}_{m'}(t)])\langle D^\dag (t)D  (\tau)\rangle\nonumber\\
&&+t\leftrightarrow\tau
]\nonumber\\
&-& h_{sm,s'm'}[(\Re[\vec{\tilde{K}}_m(t)\cdot \vec{\tilde{K}}_m(\tau)] +is \Re[\vec{M}^z_m(t,\tau)])\langle D(t)D(\tau)\rangle\nonumber\\
&+&(\Re[\vec{\tilde{K}}^*_m(t)\cdot \vec{\tilde{K}}_m(\tau)] +is \Re[\vec{N}^z_m(t,\tau)])\langle D^\dag(t)D(\tau)\rangle]\nonumber\\
&-& h_{\bar{s}m,s'm'}[(s\Re[\vec{M}^y_m(t,\tau)] +i\Re[\vec{M}^x_m(t,\tau)]) \langle D(t)D(\tau)\rangle\nonumber\\
&+&(s\Re[\vec{N}^y_m(t,\tau)] +i\Re[\vec{N}^x_m(t,\tau)])  \langle D^\dag(t)D(\tau)\rangle]\nonumber\\
&-&h_{sm,s'm'}[(\Re[\vec{\tilde{K}}_{m'}(t)\cdot \vec{\tilde{K}}_{m'}(\tau)] -is' \Re[\vec{M}^z_{m'}(t,\tau)])\langle D(\tau)D(t)\rangle\nonumber\\
&+&(\Re[\vec{\tilde{K}}_{m'}(t)\cdot \vec{\tilde{K}}^*_{m'}(\tau)] -is' \Re[\vec{N}^{*z}_{m'}(t,\tau)])\langle D^\dag(\tau)D(t)\rangle]\nonumber\\
&-&h_{sm,\bar{s}'m'}[(s'\Re[\vec{M}^y_{m'}(t,\tau)] -i\Re[\vec{M}^x_{m'}(t,\tau)]) \langle D(\tau)D(t)\rangle\nonumber\\
&+&(s'\Re[\vec{N}^{*y}_{m'}(t,\tau)] -i\Re[\vec{N}^{*x}_{m'}(t,\tau)])  \langle D^\dag(\tau)D(t)\rangle]\},
\end{eqnarray}
where $\vec{M}_m(t,\tau)=\vec{\tilde{K}}_m(t)\times \vec{\tilde{K}}_m(\tau)$, and $\vec{N}_m(t,\tau)=\vec{\tilde{K}}^*_m(t)\times \vec{\tilde{K}}_m(\tau)$ with $\vec{\tilde{K}}_m(t)=( \tilde{K}^x_m(t),\tilde{K}^y_m(t),\tilde{K}^z_m(t))$.
\par The first order inhomogeneous terms are
\begin{eqnarray}
[\dot{h}_{1m,1m'}]_{ih1}=0,
\end{eqnarray}
\begin{eqnarray}
&&[\dot{h}_{1m,\bar{1}m'}]_{ih1}=-i{\rm Tr}_{\rm B}\{\langle 1m|[\tilde{H}_{1,I}(t),\mathcal{Q}\tilde{\rho}_{\rm I}(0)]|\bar{1}m'\rangle\}\nonumber\\
&=&-iJq_mq_{m'}([  \tilde{K}^x_m(t)-i \tilde{K}^y_m(t) ]  \langle D(t)\rangle_{Qmm'} \nonumber\\
 &&+[ \tilde{K}^{x*}_m(t)-i \tilde{K}^{y*}_m(t) ]  \langle D^\dag(t)\rangle_{Qmm'} ),
\end{eqnarray}
\begin{eqnarray}
&&[\dot{h}_{\bar{1}m,1m'}]_{ih1}=-i{\rm Tr}_{\rm B}\{\langle \bar{1}m|[\tilde{H}_{1,I}(t),\mathcal{Q}\tilde{\rho}_{\rm I}(0)]|1m'\rangle\}\nonumber\\
&=&iJq_mq_{m'}( [  \tilde{K}^x_{m'}(t)+i \tilde{K}^y_{m'}(t)  ]  \langle D(t)\rangle_{Qmm'} \nonumber\\
&&+ [  \tilde{K}^{x*}_{m'}(t)+i \tilde{K}^{y*}_{m'}(t)  ]  \langle D^\dag(t)\rangle_{Qmm'}),
\end{eqnarray}
and
\begin{eqnarray}
&&[\dot{h}_{\bar{1}m,\bar{1}m'}]_{ih1}=-i{\rm Tr}_{\rm B}\{\langle \bar{1}m|[\tilde{H}_{1,I}(t),\mathcal{Q}\tilde{\rho}_{\rm I}(0)]|\bar{1}m'\rangle\}\nonumber\\
&=&iJq_mq_{m'}([\tilde{K}^z_m(t) -\tilde{K}^z_{m'}(t)]  \langle D(t)\rangle_{Qmm'} \nonumber\\
 &&+[ \tilde{K}^{z*}_m(t)-\tilde{K}^{z*}_{m'}(t)  ]  \langle D^\dag(t)\rangle_{Qmm'}),
\end{eqnarray}
where $\langle D(t)\rangle_{Qmm'}={\rm Tr}_{\rm B}\{D(t)\delta\rho^{mm'}_{\rm B} \}$.
Direct calculation gives
\begin{eqnarray}
&&\langle D(t)\rangle_{Qmm'}=\Theta e^{-\frac{1}{2}(m-m')^2\psi_1(0)}\nonumber\\
&&\{e^{-\psi_{mm'}(t)}[(d_m(t)+1)(d_{m'}(t)+1)]^{\frac{1}{2}}-1\},
\end{eqnarray}
and $\langle D^\dag(t)\rangle_{Qmm'}=\langle D(t)\rangle_{Qm'm}^*$.
\par The second order inhomogeneous terms are
\begin{eqnarray}
&&[ \dot{h}_{1m,1m'} ]_{ih2}=q_mq_{m'}J^2\int^t_0d\tau\nonumber\\
&&\{ (\tilde{K}^x_m(\tau) -i \tilde{K}^y_m(\tau)) (\tilde{K}^x_{m'}(t) +i \tilde{K}^y_{m'}(t))   \langle D(t)D(\tau)\rangle_{Qmm'}\nonumber\\
&&+  (\tilde{K}^{x*}_m(\tau) -i \tilde{K}^{y*}_m(\tau)) (\tilde{K}^x_{m'}(t) +i \tilde{K}^y_{m'}(t)) \langle D(t)D^\dag(\tau)\rangle_{Qmm'}\nonumber\\
&&+ (\tilde{K}^x_m(\tau) -i \tilde{K}^y_m(\tau))  (\tilde{K}^{x*}_{m'}(t) +i \tilde{K}^{y*}_{m'}(t))  \langle D^\dag(t)D(\tau)\rangle_{Qmm'}\nonumber\\
&&+  (\tilde{K}^{x*}_m(\tau) -i \tilde{K}^{y*}_m(\tau)) (\tilde{K}^{x*}_{m'}(t) +i \tilde{K}^{y*}_{m'}(t)) \langle D^\dag(t)D^\dag(\tau)\rangle_{Qmm'}\nonumber\\
&& +t\leftrightarrow\tau\},
\end{eqnarray}
\begin{eqnarray}
&&[ \dot{h}_{1m,\bar{1}m'} ]_{ih2}=-J^2q_mq_{m'}\int^t_0d\tau\nonumber\\
&&\{[ (\tilde{K}^x_m(\tau) -i \tilde{K}^y_m(\tau))  \tilde{K}^z_{m'}(t)  \langle D(t)D(\tau)\rangle_{Qmm'}\nonumber\\
 &&+  (\tilde{K}^{x*}_m(\tau) -i \tilde{K}^{y*}_m(\tau))  \tilde{K}^z_{m'}(t)   \langle D(t)D^\dag(\tau)\rangle_{Qmm'}\nonumber\\
&&+ (\tilde{K}^x_m(\tau) -i \tilde{K}^y_m(\tau))  \tilde{K}^{z*}_{m'}(t)  \langle D^\dag(t)D(\tau)\rangle_{Qmm'} \nonumber\\
&&+  (\tilde{K}^{x*}_m(\tau) -i \tilde{K}^{y*}_m(\tau)) \tilde{K}^{z*}_{m'}(t)   \langle D^\dag(t)D^\dag(\tau)\rangle_{Qmm'} \nonumber\\
&&+t\leftrightarrow\tau]\nonumber\\
&&+  [ \vec{M}^y_m(t,\tau) +i\vec{M}^x_m(t,\tau)]  \langle D(t)D(\tau)\rangle_{Qmm'}\nonumber\\
&& + [ \vec{N}^y_m(t,\tau) +i\vec{N}^x_m(t,\tau)]   \langle D^\dag(t)D(\tau)\rangle_{Qmm'}\nonumber\\
&&+  [ \vec{N}^{*y}_m(t,\tau) +i\vec{N}^{*x}_m(t,\tau)]  \langle D(t)D^\dag(\tau)\rangle_{Qmm'} \nonumber\\
&&+  [ \vec{M}^{*y}_m(t,\tau) +i\vec{M}^{*x}_m(t,\tau)]   \langle D^\dag(t)D^\dag(\tau)\rangle_{Qmm'} \},
\end{eqnarray}
\begin{eqnarray}
&&[ \dot{h}_{\bar{1}m,1m'} ]_{ih2}=- q_mq_{m'}J^2\int^t_0d\tau\nonumber\\
&&\{  [\tilde{K}^z_m(\tau)   (\tilde{K}^x_{m'}(t) +i \tilde{K}^y_{m'}(t)) \langle D(t)D(\tau)\rangle_{Qmm'}\nonumber\\
&& +   \tilde{K}^{z*}_m(\tau)  (\tilde{K}^x_{m'}(t) +i \tilde{K}^y_{m'}(t)) \langle D(t)D^\dag(\tau)\rangle_{Qmm'}\nonumber\\
&&+  \tilde{K}^z_m(\tau)  (\tilde{K}^{x*}_{m'}(t) +i \tilde{K}^{y*}_{m'}(t))  \langle D^\dag(t)D(\tau)\rangle_{Qmm'}\nonumber\\
&& +  \tilde{K}^{z*}_m(\tau)  (\tilde{K}^{x*}_{m'}(t) +i \tilde{K}^{y*}_{m'}(t))  \langle D^\dag(t)D^\dag(\tau)\rangle_{Qmm'}  \nonumber\\
&&+t\leftrightarrow\tau ]\nonumber\\
&&+[ \vec{M}^y_{m'}(t,\tau) -i\vec{M}^x_{m'}(t,\tau)]   \langle D(\tau)D(t)\rangle_{Qmm'}\nonumber\\
&& +   [ \vec{N}^{*y}_{m'}(t,\tau) -i\vec{N}^{*x}_{m'}(t,\tau)] \langle D^\dag(\tau)D(t)\rangle_{Qmm'}\nonumber\\
&&+  [ \vec{N}^{y}_{m'}(t,\tau) -i\vec{N}^{x}_{m'}(t,\tau)]  \langle D(\tau)D^\dag(t)\rangle_{Qmm'}\nonumber\\
&& +   [ \vec{M}^{*y}_{m'}(t,\tau) -i\vec{M}^{*x}_{m'}(t,\tau)]  \langle D^\dag(\tau)D^\dag(t)\rangle_{Qmm'}\},
\end{eqnarray}
and
\begin{eqnarray}
&&[\dot{h}_{\bar{1}m,\bar{1}m'}]_{ih2}=q_mq_{m'}J^2\int^t_0d\tau\nonumber\\
&=&\{ [\tilde{K}^z_m(\tau) \tilde{K}^z_{m'}(t)  \langle D(t)D(\tau)\rangle_{Qmm'}\nonumber\\
&& + \tilde{K}^{z*}_m(\tau) \tilde{K}^z_{m'}(t)  \langle D(t)D^\dag(\tau)\rangle_{Qmm'}\nonumber\\
&+& \tilde{K}^z_m(\tau) \tilde{K}^{z*}_{m'}(t)  \langle D^\dag(t)D(\tau)\rangle_{Qmm'}\nonumber\\
&& +  \tilde{K}^{z*}_m(\tau) \tilde{K}^{z*}_{m'}(t) \langle D^\dag(t)D^\dag(\tau)\rangle_{Qmm'} \nonumber\\
&&+t\leftrightarrow\tau]\nonumber\\
&-&(  [\vec{\tilde{K}}_m(t)\cdot \vec{\tilde{K}}_m(s) -i \vec{M}^z_m(t,s) ]
  \langle D(t)D(s)\rangle_{Qmm'} \nonumber\\
&&+  [\vec{\tilde{K}}^*_m(t)\cdot \vec{\tilde{K}}_m(s) -i  \vec{N}^z_m(t,s) ]
  \langle D^\dag(t)D(s)\rangle_{Qmm'}\nonumber\\
&&+  [\vec{\tilde{K}}_m(t)\cdot \vec{\tilde{K}}^*_m(s) -i  \vec{N}^{*z}_m(t,s) ]
  \langle D(t)D^\dag(s)\rangle_{Qmm'}\nonumber\\
&& +  [\vec{\tilde{K}}^*_m(t)\cdot \vec{\tilde{K}}^*_m(s) -i  \vec{M}^{*z}_m(t,s) ]
  \langle D^\dag(t)D^\dag(s)\rangle_{Qmm'}  \nonumber\\
  &&+ [\vec{\tilde{K}}_{m'}(t)\cdot \vec{\tilde{K}}_{m'}(s) +i \vec{M}^z_{m'}(t,s) ]
  \langle D(s)D(t)\rangle_{Qmm'} \nonumber\\
&&+  [\vec{\tilde{K}}_{m'}(t)\cdot \vec{\tilde{K}}^*_{m'}(s) +i \vec{N}^{*z}_{m'}(t,s) ]
   \langle D^\dag(s)D(t)\rangle_{Qmm'}\nonumber\\
&&+  [\vec{\tilde{K}}^*_{m'}(t)\cdot \vec{\tilde{K}}_{m'}(s)+i \vec{N}^{z}_{m'}(t,s) ]
 \langle D(s)D^\dag(t)\rangle_{Qmm'} \nonumber\\
&&+  [\vec{\tilde{K}}^*_{m'}(t)\cdot \vec{\tilde{K}}^*_{m'}(s) +i  \vec{M}^{*z}_{m'}(t,s) ]
 \langle D^\dag(s)D^\dag(t)\rangle_{Qmm'})\},\nonumber\\
\end{eqnarray}
where the inhomogeneous correlation functions
\begin{eqnarray}
\langle D(t)D(s)\rangle_{Qmm'}&=&Tr\{\delta^{mm'}_{\rm B}D(t)D(s)\},
\end{eqnarray}
etc., with $\delta^{mn}_{\rm B}=\tilde{\rho}_{\rm B}^{mn}-\rho_{\rm B}{\rm Tr}_{\rm B}\{\rho_{\rm B}e^{(m-n)B_1}\}$.
\par Direct calculation gives
\begin{eqnarray}
&&\langle D(t)D(s)\rangle_{Qmm'}=e^{-\frac{(m-m')^2 }{2}\psi(0)}e^{-\phi(0)}\nonumber\\
&&\{e^{-\phi(t-s)}e^{-\psi_{mm'}(t)-\psi_{mm'}(s)}[(d_m(t)+1)(d_{m'}(t)+1)\nonumber\\
&&(d_m(s)+1)(d_{m'}(s)+1)]^{\frac{1}{2}}\nonumber\\
&&-e^{-\psi_{mm'}(t)} [(d_m(t)+1)(d_{m'}(t)+1)]^{\frac{1}{2}}\nonumber\\
&&-e^{-\psi_{mm'}(s)} [(d_m(s)+1)(d_{m'}(s)+1)]^{\frac{1}{2}}\nonumber\\
&&-(e^{-\phi(t-s)}-2)\},
\end{eqnarray}
\begin{eqnarray}
&&\langle D(t)D^\dag(s)\rangle_{Qmm'}=e^{-\frac{(m-m')^2 }{2}\psi(0)}e^{-\phi(0)}\nonumber\\
&&\{e^{\phi(t-s)}e^{-\psi_{mm'}(t)+\psi_{mm'}(s)}[(d_m(t)+1)(d_{m'}(t)+1)\nonumber\\
&&(d_m(s)+1)^*(d_{m'}(s)+1)^*]^{\frac{1}{2}}\nonumber\\
&&-e^{-\psi_{mm'}(t)} [(d_m(t)+1)(d_{m'}(t)+1)]^{\frac{1}{2}}\nonumber\\
&&-e^{\psi_{mm'}(s)} [(d_m(s)+1)^*(d_{m'}(s)+1)^*]^{\frac{1}{2}}\nonumber\\
&&-(e^{\phi(t-s)}-2)\},
\end{eqnarray}
\begin{eqnarray}
&&\langle D^\dag(t)D(s)\rangle_{Qmm'}=e^{-\frac{(m-m')^2 }{2}\psi(0)}e^{-\phi(0)}\nonumber\\
&&\{e^{\phi(t-s)}e^{\psi_{mm'}(t)-\psi_{mm'}(s)}[(d_m(t)+1)^*(d_{m'}(t)+1)^*\nonumber\\
&&(d_m(s)+1)(d_{m'}(s)+1)]^{\frac{1}{2}}\nonumber\\
&&-e^{\psi_{mm'}(t)} [(d_m(t)+1)^*(d_{m'}(t)+1)^*]^{\frac{1}{2}}\nonumber\\
&&-e^{-\psi_{mm'}(s)} [(d_m(s)+1)(d_{m'}(s)+1)]^{\frac{1}{2}}\nonumber\\
&&-(e^{\phi(t-s)}-2)\},
\end{eqnarray}
and
\begin{eqnarray}
 \langle D^\dag(t)D^\dag(s)\rangle_{Qmm'}=\langle D (s)D (t)\rangle^*_{Qm'm}.
\end{eqnarray}
Then $h_{sm,s'm'}$ are obtained by integrating the right hand side of Eq.(\ref{hEOM}) numerically.
\subsection{Calculation of $[\Theta^{(\frac{N}{2})}]^Q_{mn}$ in Eq.(\ref{ThetaQ})}
The irrelevant contribution can be calculated directly:
\begin{eqnarray}
&&[\Theta^{(\frac{N}{2})}_{\rm L}]^Q_{mn}\nonumber\\
&=&{\rm Tr}_{\rm S+B}\{ \langle m| \mathcal{ Q} \tilde{\rho}_{\rm I}(0)|n\rangle e^{i(H_n-H_m)t} e^{(n-m)B_1(t)}\}\nonumber\\
&=&q_mq_ne^{-(m-n)^2\psi_1(0)}\cdot\nonumber\\
&&\left(\sin^2\frac{\theta_{nm}}{2}e^{iE_{nm,+}t}+\cos^2\frac{\theta_{nm}}{2}e^{iE_{nm,-}t}\right)\nonumber\\
&&(e^{(n-m)[(n-m)\psi_1(t)+i(n+m)\psi_2(t)-i\sum_k\frac{\eta_k\xi_k}{\omega^2_k}\sin\omega_kt]}-1).\nonumber\\
\end{eqnarray}

\end{document}